\documentclass[twocolumn]{aastex701}

\usepackage{multirow}
\usepackage{aas_macros}
\usepackage{verbatim}
\usepackage{adjustbox}
\usepackage{xcolor}
\usepackage[flushleft]{threeparttable}
\usepackage[utf8]{inputenc}
\usepackage{amsmath}	
\usepackage{amssymb}	
\usepackage[graphicx]{realboxes}
\usepackage{afterpage}
\usepackage{longtable}
\usepackage{supertabular}
\usepackage{comment}
\usepackage{float}
\usepackage{verbatim}
\usepackage{threeparttable}
\usepackage{latexsym}
\usepackage{epsf}
\usepackage{caption}
\usepackage[titletoc,title]{appendix}
\include{journals}

\makeatletter
\def\orcid#1{}
\def\@orcid{}

\makeatother

\begin{document}

\title{Study of Integrated Far-ultraviolet Emissions from Galactic Globular Clusters using AstroSat/UVIT observations}

\correspondingauthor{Ananta C. Pradhan}
\email{acp.phy@gmail.com; sonika6195@gmail.com}

\author{Sonika Piridi}
\affiliation{Department of Physics and Astronomy, National Institute of Technology, Rourkela 769008, India}
\email{sonika6195@gmail.com}

 \author{Ranjan Kumar}
 \affiliation{ Department of Physics, U. R. College, Rosera, Samastipur, A constituent unit of Lalit Narayan Mithila University, Darbhanga, Bihar - 848210, India}
 \email{ranjankmr488@gmail.com}
 
 \author{Divya Pandey}
 \affiliation{Tartu Observatory, University of Tartu, Observatooriumi 1, 61602 Tõravere, Estonia}
 \email{divyapandey1212@gmail.com}
 
 \author{Ananta C. Pradhan}
 \affiliation{Department of Physics and Astronomy, National Institute of Technology, Rourkela 769008, India}
 \email{acp.phy@gmail.com}

\begin{abstract}
We used observations obtained with the Ultraviolet Imaging Telescope on board the AstroSat satellite to measure the integrated far-ultraviolet (FUV) and optical (V) magnitudes of 30 Galactic globular clusters (GCs). We classified the UV-bright evolved stellar populations of the GCs using FUV$-$V versus FUV color-magnitude diagrams (CMDs) and BaSTI-IAC isochrones and subsequently quantified their contributions to the total integrated FUV emissions. We found that the horizontal branch (HB) and post-HB (post-HB) stars contribute $\sim 40\%-45$\% to the total FUV emission of GCs, while the contribution of blue straggler stars is only $\sim$3\%. The HB stars especially dominate the UV budget of the metal-poor clusters. The observed spread in FUV-optical color in the color-color diagram supports the phenomenon that the UV upturn of early-type galaxies is due to the evolved stars. We studied for the first time the variation of integrated FUV magnitudes and colors with several cluster parameters in the core, intermediate, outer, and tidal regions, such as the fraction of second-generation stars, helium mass fraction, HB morphology, and mass of the GCs. We found that the GCs with a higher second-generation star fraction, helium mass fraction, and cluster mass are brighter in all the regions. The GCs with bluer HB morphologies also have brighter and bluer FUV magnitudes in the core and intermediate regions. Metal-poor GCs show significantly bluer FUV$-$optical colors, consistent with a stronger contribution from hot evolved stars. 
\end{abstract}

\keywords{\uat{Globular star clusters}{656} --- \uat{Ultraviolet astronomy}{1736} --- \uat{Integrated magnitude}{800} --- \uat{Stellar astronomy}{1583} --- \uat{Horizontal branch stars}{746}}


\section{Introduction}
Globular clusters (GCs) are gravitationally bound systems comprising old stellar populations, typically residing in the halo or thick disk of the galaxy. They are mainly composed of older stars, including horizontal branch (HB), extreme HB (EHB), post-HB (post-HB), and asymptotic giant branch (AGB) stars. They also contain some exotic stars such as blue stragglers (BSs), cataclysmic variables (CVs), and white dwarfs (WDs). Although these stars are significantly fewer in number than the faint low-mass main-sequence (MS) stars, they represent the primary source of ultraviolet (UV) light in the old stellar populations of GCs \citep{2001moehler_hotstars, 2009catelan_hb, Dalessandro_2012, Schiavon2012}. They are more easily detectable in UV images, offering valuable insights into their properties that may be missed at optical wavelengths. The  UV integrated magnitudes and colors of Galactic GCs are predominantly shaped by these evolved, old, and hot stellar populations.

The  {\em Galaxy Evolution Explorer (GALEX)} and the {\em Hubble Space Telescope (HST)} are pivotal UV telescopes that have significantly advanced our understanding of GCs by analyzing their UV emissions \citep{Dalessandro_2012, 2018hugs}. Recently, \citet{Dalessandro_2012} examined the integrated UV colors of 44 GCs using {\em GALEX} data, revealing that these colors predominantly originate from hot stellar populations, particularly HB stars and their progeny. Their findings indicate strong correlations between UV colors and key GC parameters, such as metallicity, age, helium abundance, concentration, and the HB morphology ratio (HBR), which collectively influence the morphology of the HB. Additional UV telescopes, including the {\em Ultraviolet Imaging Telescope} \citep[UIT;][]{whitney1994far}, and the {\em International Ultraviolet Explorer} \citep[IUE;][]{castellani1987globular}, have further enriched our knowledge of GCs. These instruments have provided critical insights into the integrated UV magnitudes and their connections to the physical properties of GCs, enhancing our understanding of their stellar populations and evolutionary dynamics. Recently, the availability of numerous multiwavelength observations in the infrared, optical, and UV bands has enabled the identification of stars at various evolutionary stages through their locations on the color-magnitude diagram (CMD). Furthermore, Gaia EDR3 proper-motion data have facilitated the separation of cluster members from field stars \citep{gaia-gc}.

In distant extragalactic studies, we often depend on the integrated magnitudes of GCs because resolving individual stars within these clusters is difficult due to the limited resolving power of telescopes \citep{Dalessandro_2012}. By analyzing the integrated magnitudes of Galactic GCs and their stellar populations, we can infer the properties of stellar constituents in extragalactic GCs. Analysis of the UV flux distribution within GCs provides valuable insight into their formation history, including factors such as age, variations in intracluster metallicity through multiwavelength observations, and the helium abundance of the stellar population \citep{Dalessandro_2012, 2018peacock_hb}. Furthermore, the UV integrated light from GCs helps to understand the ``UV upturn'' phenomenon observed in early-type galaxies (ETGs), characterized by an increase in UV flux at shorter wavelengths, primarily driven by hot, evolved stellar populations \citep{1995dorman, 2004yoon_uvupturn, 2007kaviraj, 2009peng_uvupturn, 2016Schombert, 2024krishna_uvupturn}.

The Ultraviolet Imaging Telescope (UVIT) on board AstroSat has observed many GCs in far-ultraviolet (FUV) and near-ultraviolet (NUV) wavebands. With an angular resolution better than 1.5$''$ and a 28$'$ diameter field of view (FoV), UVIT enables high-precision studies of UV-bright stellar populations within GCs \citep{2017Tandonfirst, 2020tandon}. Its exceptional resolution facilitates the analysis of individual stars, providing insights into their properties through multiwavelength spectral energy distributions (SEDs). These SEDs allow researchers to determine key stellar parameters such as mass, luminosity, and effective temperature \citep{2017ngc1851hbstars, 2019sahu_bss_wd,  Singh2020,  ranjan7492,  ranjanngc4590,m3m13,2024e3, 2023panthi}. Recently, \cite{sneha8cluster} have studied the FUV$-$optical CMDs of 11 GCs observed with UVIT, revealing a correlation between the number of EHB stars detected in FUV filters and the maximum helium variation within GCs. These results underscore the critical role of UVIT in advancing our understanding of the stellar populations and evolutionary processes in GCs, leveraging its high-resolution imaging.

In this work, we studied 30 GCs observed by UVIT. We segmented the structure of GCs into three different regions: the core region (denoted by $\delta_{c}$), the region between the core and the half-light radius (denoted by $\delta_{hc}$), and the region between the half-light and the tidal radius (denoted by $\delta_{th}$). We estimated the integrated UV magnitudes and colors of GCs observed by UVIT across these regions. We also investigated the influence of helium abundance, the fraction of multiple stellar populations, the HBR, and other factors on the UV magnitudes and UV$-$optical colors of GCs. All the UVIT magnitudes used throughout the paper are in the AB magnitude system.

\section{UVIT observations of globular clusters} \label{sec:data}
\subsection{Data Analysis}

\begin{figure}[ht]
    \centering
   \includegraphics[width=\columnwidth]{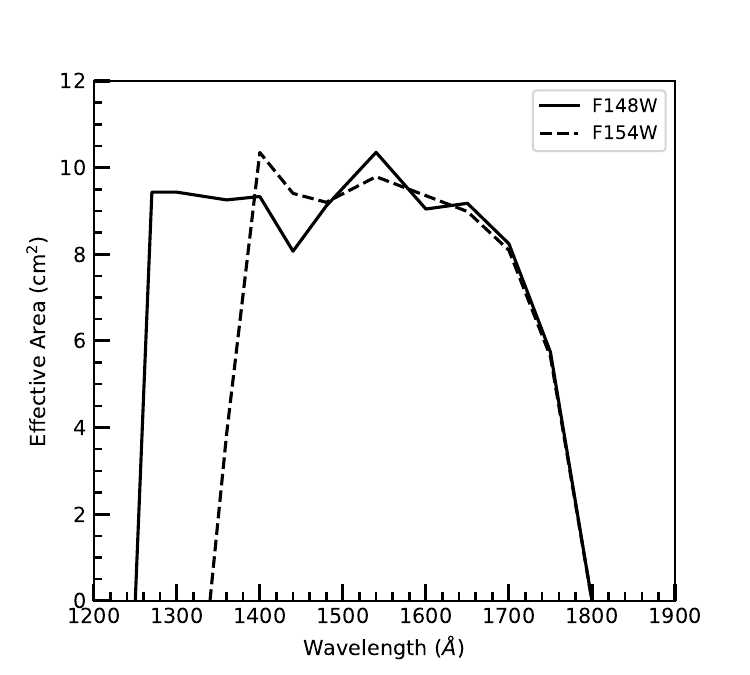}
     \caption{ Filter response curves of the two UVIT FUV filters, F148W and F154W. We used a sample of a GCs observed with these filters.} 
    \label{fig:filter-response}
\end{figure}

\begin{deluxetable*}{|l|l|c|r|r|r|r|r|r|r|r|}
\tablecaption{ Details of Galactic GCs observed by UVIT, along with their exposure times, FUV and optical (V) surface brightness in the core, intermediate, outer, and tidal regions.\label{tab:gc-mag}}
\tablehead{Cluster Name &Filter  & FUV Exp (s)&$\rm FUV_{c}$ &$\rm FUV_{hc}$ &$\rm FUV_{th}$ &$\rm FUV_{t}$ &$\rm V_{c}$ &$\rm V_{hc}$&$\rm V_{th}$&$\rm V_{t}$
  }

\startdata
  NGC 104 & F148W & 12249 & 10.24 & 11.97 & $-$ & $-$ & 1.55 & 4.33 & 10.73 & 9.59\\
  NGC 6637 & F148W & 3891 & 9.58 & 11.11 & 12.63 & 12.58 & 2.55 & 3.75 & 8.64 & 7.79\\
  E3 & F154W & 1846 & 8.93 & 10.68 & 14.99 & 12.52 & 8.96 & 9.63 & 13.76 & 12.28\\
  NGC 1851 & F148W & 6654 & 5.33 & 6.62 & 14.30 & 12.97 & $-$0.92 & 1.50 & 7.94 & 7.24\\
  NGC 6723 & F148W & 6583 & 8.81 & 9.95 & 12.46 & 12.12 & 3.81 & 5.01 & 8.99 & 7.98\\
  NGC 2808 & F154W & 4988 & 4.47 & 5.27 & 11.00 & 10.51 & 0.37 & 1.77 & 7.83 & 7.15\\
  NGC 362 & F148W & 4285 & 7.00 & 9.16 & $-$ & $-$ & 0.62 & 2.40 & 8.78 & 7.97\\
  NGC 288 & F148W & 6229 & 9.91 & 10.81 & 13.95 & 13.34 & 5.81 & 6.19 & 9.78 & 9.09\\
  NGC 6981 & F148W & 5142 & 8.99 & 10.61 & 12.75 & 12.61 & 3.29 & 4.38 & 9.00 & 8.21\\
  NGC 3201 & F148W & 6706 & 10.94 & 11.94 & $-$ & $-$ & 3.97 & 6.13 & 12.60 & 10.67\\
  M3 & F148W & 2988 & 7.50 & 9.68 & $-$ & $-$ & 1.12 & 3.42 & 10.64 & 9.05\\
  NGC 6584 & F154W & 1148 & 9.94 & 9.87 & 13.27 & 13.14 & 2.59 & 3.78 & 8.47 & 7.99\\
  NGC 5139 & F148W & 6182 & 7.93 & 8.75 & $-$ & $-$ & 4.56 & 4.95 & 9.58 & 9.19\\
  IC 4499 & F148W & 2626 & 10.40 & 11.15 & 13.81 & 13.56 & 5.31 & 6.05 & 10.10 & 9.36\\
  NGC 6205 & F148W & 6646 & 7.08 & 7.88 & $-$ & $-$ & 2.95 & 3.89 & 9.50 & 8.80\\
  NGC 5986 & F148W & 7023 & 6.54 & 7.99 & 11.96 & 11.49 & 2.78 & 3.69 & 8.74 & 7.90\\
  NGC 1904 & F148W & 4696 & 5.43 & 6.75 & 11.98 & 11.31 & 2.00 & 3.37 & 8.05 & 7.6\\
  NGC 7492 & F154W & 356 & 9.72 & 10.20 & 11.93 & 11.82 & 5.44 & 6.12 & 9.78 & 9.14\\
  NGC 4147 & F154W & 1653 & 5.91 & 7.92 & 12.81 & 12.29 & 2.14 & 3.84 & 9.04 & 8.41\\
  NGC 6809 & F148W & 6590 & 9.88 & 10.63 & $-$ & $-$ & 5.48 & 6.30 & 9.68 & 8.95\\
  NGC 5466 & F148W & 4849 & 11.12 & 11.95 & $-$ & $-$ & 5.66 & 6.33 & 11.81 & 11.09\\
  NGC 2298 & F148W & 2301 & 8.08 & 9.60 & 12.43 & 12.08 & 3.94 & 5.37 & 9.23 & 8.48\\
  NGC 5897 & F148W & 12691 & 9.99 & 10.48 & 12.78 & 12.46 & 5.68 & 6.67 & 9.57 & 8.83\\
  NGC 6397 & F148W & 3790 & 7.95 & 11.21 & $-$ & $-$ & 4.40 & 7.20 & 10.13 & 9.82\\
  NGC 6101 & F154W & 746 & 9.49 & 10.48 & 11.75 & 11.61 & 4.73 & 4.99 & 7.79 & 7.49\\
  NGC 2419 & F148W & 1582 & 4.79 & 5.80 & 8.09 & 8.00 & 1.23 & 2.36 & 6.94 & 6.27\\
  NGC 7099 & F148W & 6553 & 5.23 & 8.50 & $-$ & $-$ & 1.41 & 3.80 & 10.12 & 9.32\\
  NGC 5053 & F148W & 173 & 11.98 & 12.97 & 13.98 & 13.83 & 6.49 & 7.09 & 10.57 & 9.67\\
  NGC 4590 & F154W & 539 & 6.02 & 7.65 & $-$ & $-$ & 3.67 & 4.85 & 10.89 & 10.17\\
  NGC 6341 & F148W & 13922 & 9.01 & 10.64 & $-$ & $-$ & 1.31 & 3.00 & 8.72 & 7.92\\
\enddata
\tablecomments{$\rm FUV_{c} (V_{c})$, $\rm FUV_{hc}(V_{hc})$, $\rm FUV_{th}(V_{th})$ and $\rm FUV_{t}(V_{t})$ are the FUV (optical) surface brightness in the core region, intermediate region (between core and half-light radius), outer region (between half-light and tidal radius), and tidal region of the GC, respectively. The outer and total tidal surface brightness values are not available for GCs having a tidal radius greater than the UVIT FoV, denoted by `$-$'.}
\end{deluxetable*}

\begin{figure*}[ht]
    \centering   
     \includegraphics[width=\textwidth]{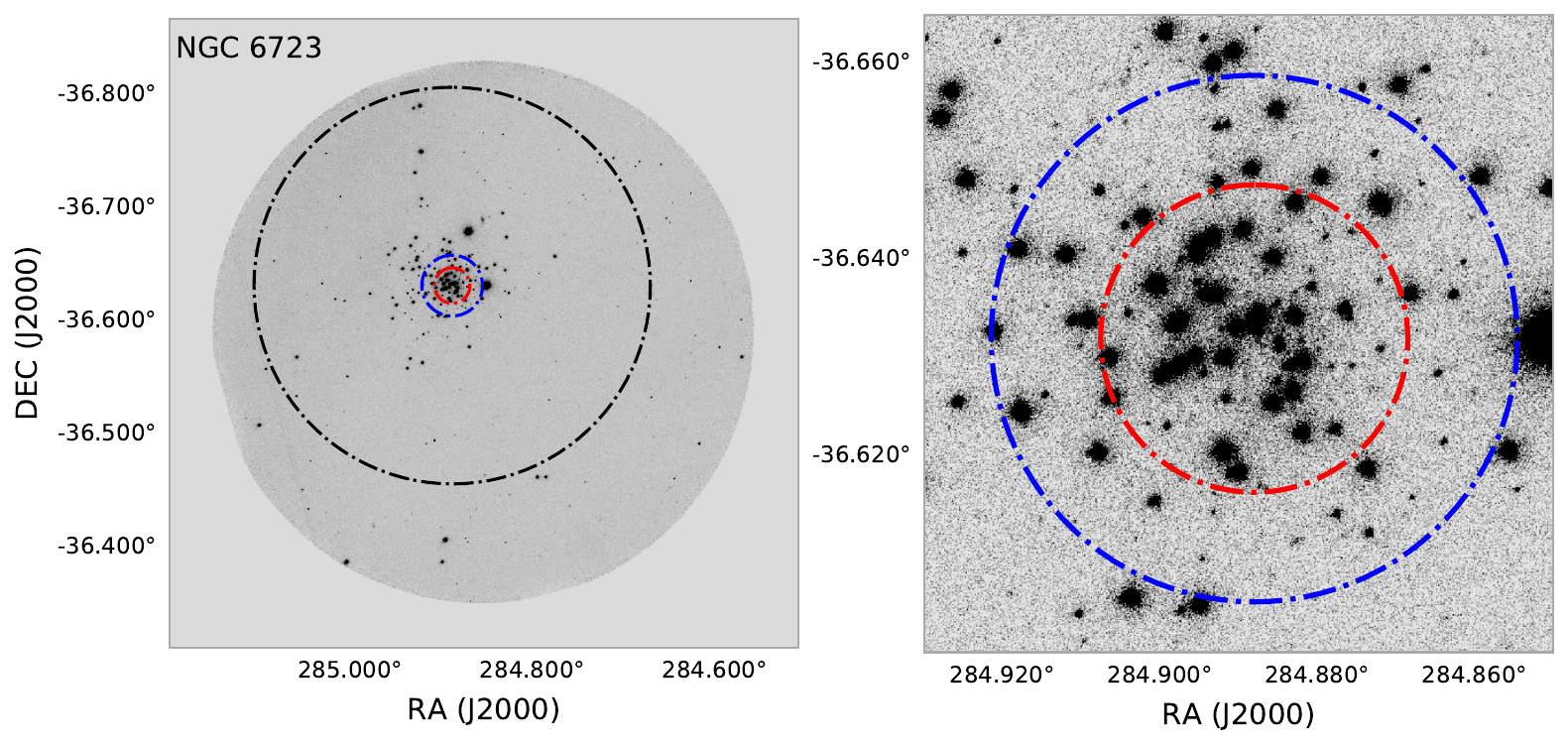}
\caption{Left panel: image of the GC NGC 6723 observed by the F148W filter of UVIT. The red, blue, and black circles delineate the core radius, half-light radius, and tidal radius of the cluster, respectively.
Right panel: zoomed-in view of the cluster's inner region, highlighting the core radius (red circle) and half-light radius (blue circle). } 
    \label{fig:gc-example}
\end{figure*}

We analyzed 30 publicly available GCs observed by UVIT. The sample comprises 11 GCs from the UVIT DR1 catalog \citep{2024uvitdr1}, 17 GCs from the UVIT DR2 catalog (S. Piridi et al. 2026, in preparation), which covers UVIT observations up to 2021, and two additional GCs observed in 2022. The effective areas and the wavelength coverages of the two filters, F148W (CaF2) and F154W (BaF2) are almost similar as shown in \autoref{fig:filter-response}. Hence, to estimate the FUV integrated magnitudes, we used the F148W filter for 23 GCs and the F154W filter for the remaining seven GCs, as FUV observations were not available in a single filter for all GCs. The NUV observations were available for only a few GCs and were therefore excluded from the analysis. The selected GCs and their corresponding UVIT filters, along with the respective exposure times, are provided in \autoref{tab:gc-mag}. We display the FUV image of a GC NGC 6723 in \autoref{fig:gc-example}. The left panel of the figure  displays the core radius ($r_{c}$), half-light radius ($r_{h}$), and tidal radius ($r_{t}$) indicated by red, blue, and black circles, respectively, whereas the right panel shows a magnified view of the cluster's core and half-light region. These radii were obtained from the updated catalog of \cite{harris2010new}.

We determined the background of UVIT images by placing 21 $\times$ 21 pixel boxes in source-free regions and fitted the Gaussian curves to obtain the mean ($\mu$) and standard deviation ($\sigma$) for each box. The overall background mean for the entire image was computed as the average of the background means from all boxes. We used source-extractor \citep[SE;][]{sextractor} for source detection with a detection threshold of 3$\sigma$ and obtained the Kron magnitudes \citep{Kron1980} for the detected sources. The detailed photometric procedures are mentioned in \cite{2024uvitdr1}. 


\subsection{Completeness}
We used the artificial star test to find the completeness limits of the GC photometry catalog \citep{2019souradeep-comp, 2024completeness-sipra}, where we injected artificial stars on the science image and tried to recover the injected stars. To inject the stars, we created a fake catalog based on the radial density profile and the luminosity function of the photometric catalog. We then injected the stars in the intermediate and outer regions of the GC, separately. To avoid artificially increasing crowding in the image, we added 5\% of the total number of real sources at a time from our fake catalog in a 0.5 mag bin. The ADDSTAR routine in IRAF utilizes the point-spread function of the image to inject these stars on the image and generates a new image with the fake stars. We followed the same source detection and the photometry process on this new image to generate our initial catalog. The ratio of recovered stars to the injected stars gives the recovery fraction or the completeness limit of the image. 

\begin{figure}
    \centering
    \includegraphics[width=\linewidth]{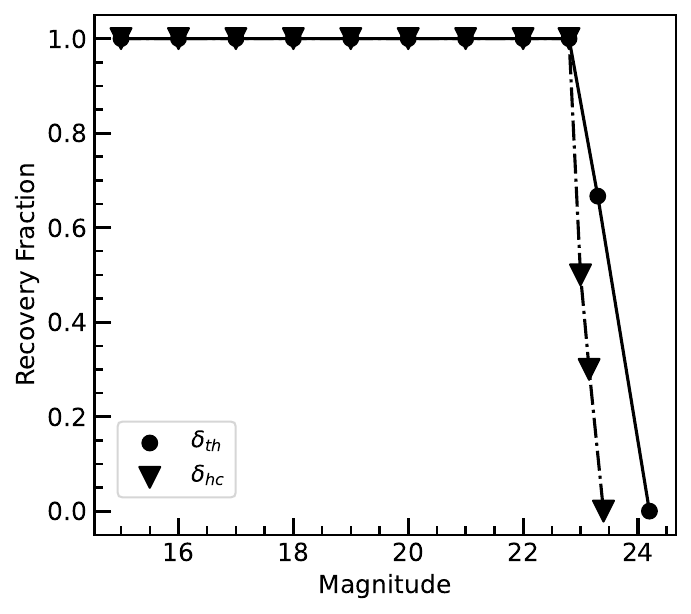}
    \caption{Recovery fraction of the artificial stars injected at each magnitude bin in the intermediate (solid downward triangles) and outer (solid circles) regions.}
    \label{fig:comp}
\end{figure}

In all the regions, we found that the real photometry catalog obtained using our source-extractor parameters is 100\% complete up to the faintest detected magnitude in the catalog i.e., we could recover all the sources from our real catalog. Hence, we tried to inject and recover the fainter stars in the intermediate and outer regions. In the intermediate region of the cluster NGC 6723 observed in the F148W filter, we found the 100\% and 50\% completeness limit to be at 22.8 and 23.5 mag, respectively. In the tidal region, we found the 100\% and 50\% completeness limit to be at 22.8 and 23 mag, respectively. The fraction of recovered stars as a function of magnitude is shown in \autoref{fig:comp}. We did not try the artificial test in the core region of the GCs due to the crowding effect.  However, in the current work, as we were studying the integrated flux of the GC, the flux of stars missing due to the crowdedness or the de-blending effect in the core region was added during the aperture photometry process. Hence, we did not lose significant flux while computing the integrated magnitudes of the GCs.

\subsection{Membership Probability}\label{sec:membership}

 To find the integrated flux of the cluster without the contamination of field sources, it is necessary to identify the cluster members. We followed different techniques to identify the cluster members in the UV (see \autoref{sec:uv-mag}) and optical V band (see \autoref{sec:intmag-optical}). 

 In the case of the UVIT photometry catalog, we utilized the membership probabilities computed in the Gaia EDR3 catalog of GCs \citep{gaia-gc}. We crossmatched the sources from our photometry catalog using a radius of 1.5$''$ and identified field sources as those having a membership probability $\le$60\% \citep{2016gcmp60}. 

For the optical magnitudes, we used the HUGS catalog for the inner region and optical ground-based telescope (GBT) data in the outer region. The membership probabilities for the HST data are available. We selected the stars having probabilities $\ge$ 60\% as the cluster members. In the outer region, for the GBT data, we first crossmatched the GBT catalog with the Gaia EDR3 catalog of clusters and selected the stars having probabilities $\ge$ 60\% as the cluster members. The magnitude depth of the GBT catalog is around 24 mag but the Gaia EDR3 catalog in the G band has a depth of around 21 mag. Hence, toward the fainter end, GBT stars do not have a Gaia EDR3 counterpart. For such stars, we applied a magnitude cut at V $>$ 20 mag. The cluster members along the MS among this set of fainter stars were chosen using the CMD and Bag of Stellar Tracks and Isochrones\footnote{\href{http://basti-iac.oa-abruzzo.inaf.it/index.html}{http://basti-iac.oa-abruzzo.inaf.it/index.html}} \citep[BaSTI-IAC;][]{2018basti}.

\subsection{CMDs and identification of sources}

\begin{figure}[ht]
    \centering
  
     \includegraphics[width=\columnwidth]{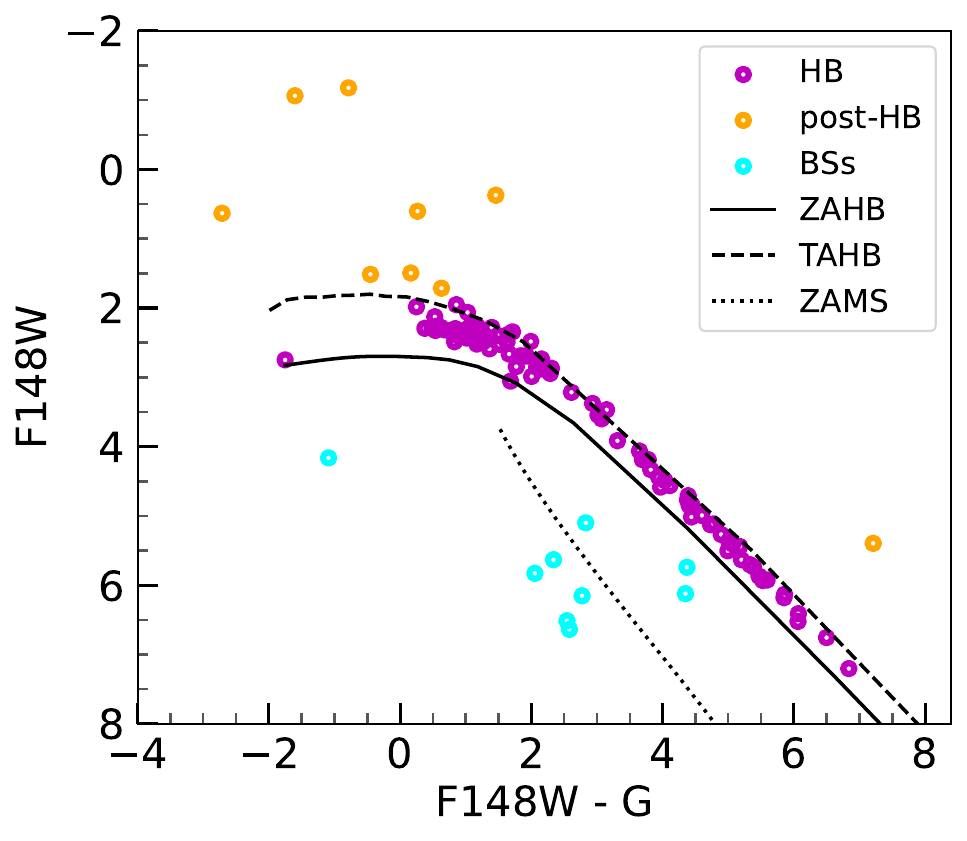}
 
\caption{F148W$-$G versus F148W CMD for one of the GC, NGC 6723. The post-HB, HB, and BS stars are shown as orange, magenta, and cyan circles, respectively. The model isochrones for ZAHB and TAHB are overplotted as solid and dashed lines.  The dotted line is the ZAMS isochrones of 0.5 Gyr.}
    \label{fig:gc-cmd}
\end{figure}

We crossmatched the UVIT sources with the HST UV Globular Cluster Survey \citep[HUGS;][]{2018hugs} for the central regions of the GCs and with the Gaia source catalog from \cite{gaia-gc} for the outer regions, as HUGS covers only a limited central portion of GCs. The HUGS and Gaia datasets provide the optical magnitudes and cluster membership probabilities for the sources. 

We classified the stellar sources of various evolutionary stages using FUV$-$optical CMDs of the GCs, following the approaches of \cite{2024uvitdr1}, and employing BaSTI-IAC \citep{2020vmag_memberprob_sep}. \autoref{fig:gc-cmd} displays F148W$–$G versus F148W CMD for one of the GCs, NGC 6723. The GCs observed in the FUV filters mainly consist of HB, post-HB, and BS stars. In the CMD, the magenta, orange, and cyan circles represent the HB, post-HB, and BS stars, respectively. The solid, dashed, and dotted lines represent the zero-age HB (ZAHB), terminal-age HB (TAHB), and zero-age MS (ZAMS) loci. \autoref{tab:gc-source-frac-region} lists the counts of the above stars present in each GC. 
 \section{Globular cluster parameters} \label{sec:cluster-param}

GC parameters provide critical diagnostics for understanding the internal structure, dynamical evolution, and formation history of GCs. We compiled these parameters for our sample from the most recent literature. The GC sample in this study spans both the northern and southern hemispheres within a galactocentric distance $R_{gc} \le$ 25 kpc, except for NGC 2419, a halo cluster at 91.5 kpc \citep{harris2010new}. The metallicities ([Fe/H]) were adopted from \citet{2014franis-feh}, who updated the values from \cite{harris2010new}. The [Fe/H] in our GC sample ranges from $-$2.31 $\le$ [Fe/H] $\le$ $-$0.69.
The age estimates of the clusters were adopted from \cite{2019mean_age} and range between 10.49 and 13.06 Gyr. The GC masses have a range of 4.06 $\rm \le \rm \log_{10} (M/M_{\odot}) \le$ 6.55 \citep{2018gcmass}. The King-model central concentration parameter (c), defined as $\rm log_{10}(r_{t}/r_{c})$, quantifies the central concentration of GCs based on the surface brightness model \citep{1966king}. The $c$ parameter ranges between 0.75 and 2.5, with 2.5 indicating the most centrally concentrated cluster. The fraction of second-generation (2G) stars (N$_{2G}$/N{$_{Tot}$) was derived from the first-generation (1G) star fractions computed by \cite{2021multiplestellarpop}. The $N_{2G}/N_{Tot}$ ranges from 0.35 to 0.91,  with approximately 70\% of the GCs having N$_{2G}$/N{$_{Tot} > 0.5$. The difference in helium abundance between 2G and 1G population stars ($\delta Y_{2G,1G}$) was sourced from \cite{milone2018hubble} and ranges within $-0.003 \le \delta Y_{2G,1G} \le$ 0.048. The HB morphology parameter (HBR) quantifies star counts on the HB, with HBR $=$1 indicating blue HB morphology and HBR $=-1$ indicating red HB morphology. The HBR values, taken from \cite{harris2010new}, range from $-1$ to 0.98 for the sample. The detailed parameters of these global properties are listed in \autoref{tab:gc-sample-param}. 

\section{Measurement of integrated magnitudes of GCs}

\subsection{Integrated FUV magnitudes} \label{sec:uv-mag}
After obtaining the science-ready UVIT images, we employed two photometry techniques to obtain the integrated FUV magnitudes of our clusters: aperture and Kron photometry. The central coordinates of the GCs were obtained from \cite{gaia-gc}. To compute the integrated magnitude using the aperture photometry technique, we first estimated the background flux per unit area by placing apertures of radius 30$''$ in source-free regions of the image using the DS9 software\footnote{\href{https://sites.google.com/cfa.harvard.edu/saoimageds9}{https://sites.google.com/cfa.harvard.edu/saoimageds9}}. This process was repeated for at least 10 different regions to calculate the mean background flux per unit area of the image. Multiplying this value by the area enclosed within the radii $r_{c}$, $r_{h}$, and $r_{t}$ yielded the total mean background flux for each region. We then measured the total fluxes within the $r_{c}$, $r_{h}$, and $r_{t}$ radii and subtracted the respective background flux to obtain the total background-subtracted flux for each region. 

The integrated flux of the field sources was calculated by summing the fluxes of individual field sources within the radii $r_{c}$, $r_{h}$, and $r_{t}$. Finally, the total integrated flux of each region was then derived by subtracting the integrated flux of the field sources from the total aperture flux obtained through the aperture photometry. We converted the integrated flux to magnitude and corrected for extinction, assuming a constant extinction within the GC, using the extinction law mentioned in \cite{cardelliext}. We used the distance moduli from \cite{gaiagcdist} to convert the magnitudes into absolute magnitudes.



Since all the clusters have different $r_{c}$, $r_{h}$, and $r_{t}$ radii, we normalized the magnitudes by log$_{10}$(area) in each region. The FUV integrated magnitudes per arcsec$^2$ or the absolute surface brightness in the $\delta_{c}$, $\delta_{hc}$, and $\delta_{th}$ regions are denoted by $\rm FUV_{c}$, $\rm FUV_{hc}$, and $\rm FUV_{th}$, respectively. Note that the UVIT FoV does not encompass the entire region up to the tidal radius for 12 of these GCs. Consequently, the $\rm FUV_{th}$ and $\rm FUV_{t}$ were not computed for these 12 GCs. The absolute surface brightness across all three regions, along with the total surface brightness ($\rm FUV_{t}$) within the tidal radius for the 30 Galactic GCs,is given in \autoref{tab:gc-mag}.

\begin{deluxetable*}{|l|c|c|c|c|c|c|c|c|}
\tablecaption{Observational details of the GC sample, along with their global parameters.\label{tab:gc-sample-param}}
\tablehead{  \multirow{2}{*}{Cluster Name} &
  \multirow{2}{*}{[Fe/H]} &
  \multicolumn{1}{c|}{Age} &
  \multicolumn{1}{c|}{Distance} &
  \multirow{2}{*}{$\rm log_{10}(M/M_{\odot})$} &
  \multirow{2}{*}{$\rm N_{2G}/N_{Tot}$} &
  \multirow{2}{*}{$\rm \delta_{Y_{2G,1G}}$} &
  \multirow{2}{*}{HBR} &
  \multirow{2}{*}{c}\\
  &  & (Gyr) & (kpc) &  &  & & &
  }

\startdata
   NGC 104 & $-$0.69 & 12.52 & 7.4 & 5.89 & 0.782 & 0.011 & $-$0.99 & 2.03\\
  NGC 6637 & $-$0.74 & 12.19 & 1.9 & 5.14 & 0.574 & 0.004 & $-$1.00 & 1.39\\
  E3 & $-$0.80 & 12.80 & 7.6 & 4.49 & $-$99.0 & $-$99.0 & $-$99.0 & 0.75\\
  NGC 1851 & $-$0.98 & 10.49 & 16.7 & 5.48 & 0.736 & 0.007 & $-$0.36 & 2.32\\
  NGC 6723 & $-$1.07 & 12.77 & 2.6 & 5.20 & 0.637 & 0.005 & $-$0.08 & 1.05\\
  NGC 2808 & $-$1.13 & 10.90 & 11.1 & 5.87 & 0.768 & 0.048 & $-$0.49 & 1.77\\
  NGC 362 & $-$1.31 & 10.87 & 9.4 & 5.54 & 0.721 & 0.008 & $-$0.87 & 1.94\\
  NGC 288 & $-$1.35 & 11.54 & 12.0 & 5.06 & 0.442 & 0.015 & 0.98 & 0.96\\
  NGC 6981 & $-$1.43 & 11.71 & 12.9 & 4.91 & 0.458 & 0.011 & 0.14 & 1.23\\
  NGC 3201 & $-$1.49 & 11.25 & 8.9 & 5.17 & 0.564 & $-$0.001 & 0.08 & 1.30\\
  NGC 6584 & $-$1.50 & 11.75 & 7.0 & 4.96 & 0.549 & 0.000 & $-$0.15 & 1.20\\
  M3 & $-$1.50 & 11.88 & 12.2 & 5.60 & 0.695 & 0.016 & 0.08 & 1.84\\
  IC 4499 & $-$1.53 & 12.00 & 15.7 & 5.22 & 0.490 & 0.004 & 0.11 & 1.11\\
  NGC 5139 & $-$1.53 & 11.52 & 6.4 & 6.55 & 0.914 & 0.033 & $-$99.0 & 1.61\\
  NGC 6205 & $-$1.58 & 12.22 & 8.7 & 5.66 & 0.816 & 0.020 & 0.97 & 1.51\\
  NGC 5986 & $-$1.63 & 12.55 & 4.8 & 5.48 & 0.754 & 0.005 & 0.97 & 1.22\\
  NGC 1904 & $-$1.66 & 11.14 & 18.8 & 5.23 & $-$99.0 & $-$99.0 & 0.89 & 1.72\\
  NGC 7492 & $-$1.72 & 12.00 & 24.9 & 4.41 & $-$99.0 & $-$99.0 & 0.81 & 1.00\\
  NGC 4147 & $-$1.81 & 12.13 & 21.3 & 4.52 & $-$99.0 & $-$99.0 & 0.55 & 1.80\\
  NGC 6809 & $-$1.86 & 12.93 & 3.9 & 5.27 & 0.689 & 0.014 & 0.87 & 0.76\\
  NGC 5466 & $-$1.98 & 13.02 & 16.2 & 4.66 & 0.533 & 0.002 & 0.58 & 1.32\\
  NGC 2298 & $-$2.03 & 12.84 & 15.7 & 4.06 & 0.630 & $-$0.003 & 0.93 & 1.28\\
  NGC 6397 & $-$2.08 & 13.06 & 6.0 & 4.95 & 0.655 & 0.006 & 0.98 & 2.50\\
  NGC 5897 & $-$2.08 & 12.30 & 7.3 & 5.31 & 0.453 & $-$99.0 & 0.86 & 0.79\\
  NGC 6101 & $-$2.10 & 12.60 & 11.1 & 5.10 & 0.346 & 0.005 & 0.84 & 0.80\\
  NGC 2419 & $-$2.15 & 12.65 & 91.5 & 5.99 & 0.630 & $-$99.0 & 0.86 & 1.40\\
  NGC 7099 & $-$2.23 & 13.06 & 7.1 & 5.12 & 0.620 & 0.015 & 0.89 & 2.50\\
  NGC 5053 & $-$2.28 & 12.68 & 16.9 & 4.75 & 0.456 & $-$0.002 & 0.52 & 0.84\\
  NGC 6341 & $-$2.31 & 13.06 & 9.6 & 5.43 & 0.696 & 0.022 & 0.91 & 1.81\\
  NGC 4590 & $-$2.31 & 12.17 & 10.1 & 5.09 & 0.619 & 0.007 & 0.17 & 1.64\\
\enddata
\tablecomments{$N_{2G}/N_{Tot}$ = fraction of 2G population, $\delta Y_{2G,1G}$ = difference in Helium abundance between 2G and 1G population, HBR = horizontal branch morphology ratio, c = central concentration. For the GCs whose $N_{2G}/N_{Tot}$, $\delta Y_{2G,1G}$, and HBR values are not available, we give the value $-$99.0.}
\end{deluxetable*}

\subsection{Integrated optical magnitudes} \label{sec:intmag-optical}
The optical V magnitudes for the sources in the inner region of the clusters were obtained from the HUGS catalog for the observed 22 GCs in our sample. HST magnitudes were converted into UBV magnitudes using the following conversion formula \citep{1995holtzman_hst_ubv, 2005sirianni_hst_ubv, 2018hsttransform}:
\begin{equation}
    y = a_{0} + a_{1}x + a_{2}x^{2}
\end{equation}
where $y$ is the difference between the BVI magnitudes and HST filters and $x$ is the intrinsic color index $(B-V)$ or $(V-I)$. The coefficients $a_{0}$, $a_{1}$, and $a_{2}$ are given in \cite{2018hsttransform}.

For sources in the outer regions of the GCs, we use data from GBT \citep{2019stetsongc}. GBT data were also used for the inner regions of eight GCs lacking HUGS data. These GCs include E3, IC 4499, NGC 1904, NGC 2419, NGC 288, NGC 4147, NGC 5139, and NGC 7492. Unlike the HUGS catalog, GBT data do not provide membership probabilities. However, we obtained the cluster membership for the GBT data as mentioned in \autoref{sec:membership}. 

We further categorized cluster members into core, intermediate, and outer regions, and calculated their integrated $V$ luminosity using the equation mentioned in \cite{baumgardt2020absolute}: 
\begin{equation}\label{eq:int-lum}
{\rm L = \Sigma_{i} 10^{-0.4V_{i}}}
 \end{equation}   
 where $V_{i}$ is the $V$ magnitude of the $i^{th}$ star.  This luminosity was converted into an integrated absolute magnitude and further transformed to the AB magnitude system  \citep{2007ab-vega-convert}. We then normalized the absolute magnitudes by log$_{10}$(area) to obtain the absolute surface brightness in each region. The optical V-band absolute surface brightness for the regions $\delta_{c}$, $\delta_{hc}$, and $\delta_{th}$ is denoted by $V_{c}$, $V_{hc}$, and $V_{th}$ is given in \autoref{tab:gc-mag}. We also provide the V band absolute surface brightness ($V_{t}$) within the total tidal radius in \autoref{tab:gc-mag}.  


\section{Results and Discussion}
\begin{deluxetable*}{|l|r|r|r|r|r|r|c|r|r|r|r|}
\tablecaption{ Flux contribution of different types of sources to the total surface brightness.\label{tab:gc-source-frac-region}}
\tablehead{ Cluster & [Fe/H] & $\rm FUV_{c}$ & $\rm FUV_{hc}$ & $\rm FUV_{th}$ & $\rm FUV_{t}$ & \rm $N_{HB}$ & \rm $N_{post-HB}$ & $\rm N_{BSs}$ & $\rm f_{HB}$ & $\rm f_{post-HB}$ & $\rm f_{BSs}$
  }

\startdata
 NGC 6637 & $-$0.74 & 9.58 & 11.11 & 12.63 & 12.58 & 1 & $-$ & 2 & 0.001 & $-$ & 0.004\\
 E3 & $-0.80$ & 8.93 & 10.68 & 14.99 & 12.52 & $-$ & 1 & $-$ & $-$ & 0.77 & $-$ \\
  NGC 1851 & $-$0.98 & 5.33 & 6.62 & 14.3 & 12.97 & 74 & 1 & 10 & 0.313 & 0.34 & 0.004\\
  NGC 6723 & $-$1.07 & 8.81 & 9.95 & 12.46 & 12.12 & 92 & 9 & 9 & 0.334 & 0.459 & 0.003\\
  NGC 2808 & $-$1.13 & 4.47 & 5.27 & 11.0 & 10.51 & 198 & 39 & 4 & 0.108 & 0.083 & 0.001\\
  NGC 288 & $-$1.35 & 9.91 & 10.81 & 13.95 & 13.34 & 119 & $-$ & 26 & 0.877 & $- $& 0.004\\
  NGC 6981 & $-$1.43 & 8.99 & 10.61 & 12.75 & 12.61 & 43 & 1 & 6 & 0.125 & 0.03 & 0.017\\
  NGC 6584 & $-$1.50 & 9.94 & 9.87 & 13.27 & 13.14 & 29 & 2 & $-$ & 0.131 & 0.021 & $-$\\
  IC 4499 & $-$1.53 & 10.39 & 11.15 & 13.81 & 13.56 & 7 & $-$ & 1 & 0.031 & $-$ & 0.011\\
  NGC 5986 & $-$1.63 & 6.54 & 7.99 & 11.96 & 11.49 & 185 & 24 & 1 & 0.269 & 0.471 & 0.001\\
  NGC 1904 & $-$1.66 & 5.43 & 6.75 & 11.98 & 11.31 & 93 & 2 & 5 & 0.268 & 0.658 & 0.002\\
  NGC 7492 & $-$1.72 & 9.72 & 10.20 & 11.93 & 11.82 & 36 & $-$ & 1 & 0.113 & $-$ & 0.003 \\ 
  NGC 4147 & $-$1.81 & 5.91 & 7.92 & 12.81 & 12.29 & 34 & 3 & 5 & 0.278 & 0.176 & 0.028\\
  NGC 2298 & $-$2.03 & 8.08 & 9.6 & 12.43 & 12.08 & 48 & 3 & 2 & 0.31 & 0.091 & 0.010\\
  NGC 5897 & $-$2.08 & 9.99 & 10.48 & 12.78 & 12.46 & 160 & 2 & 12 & 0.394 & 0.004 & 0.008\\
  NGC 6101 & $-$2.10 & 9.49 & 10.48 & 11.75 & 11.61 & 148 & $-$ & 1 & 0.415 &$ -$ & 0.001\\
  NGC 2419 & $-$2.15 & 4.79 & 5.80 & 8.09 & 8.00 &$-$ & 37 & $-$ & $-$& 0.125 & $-$\\
  NGC 5053 & $-$2.28 & 11.98 & 12.97 & 13.98 & 13.83 & 35 & $-$ & 4 & 0.153 &$-$& 0.005\\
\enddata
\tablecomments{$\rm N_{HB}$, $\rm N_{post-HB}$, and $\rm N_{BSs}$ are the number of HB, post-HB, and BS stars within the tidal radius of the GCs, respectively; $\rm f_{HB}$, $\rm f_{post-HB}$ and $\rm f_{BSs}$ represent the fractional contribution to the fluxes of the GCs by HB, post-HB and BS stars, respectively. }
\end{deluxetable*}

\subsection{Contribution of various evolved stars to the total integrated FUV emission} \label{sec:fuv-emission-hb-post-hb}

\begin{figure*}
    \centering
     
      \includegraphics[width=\textwidth]{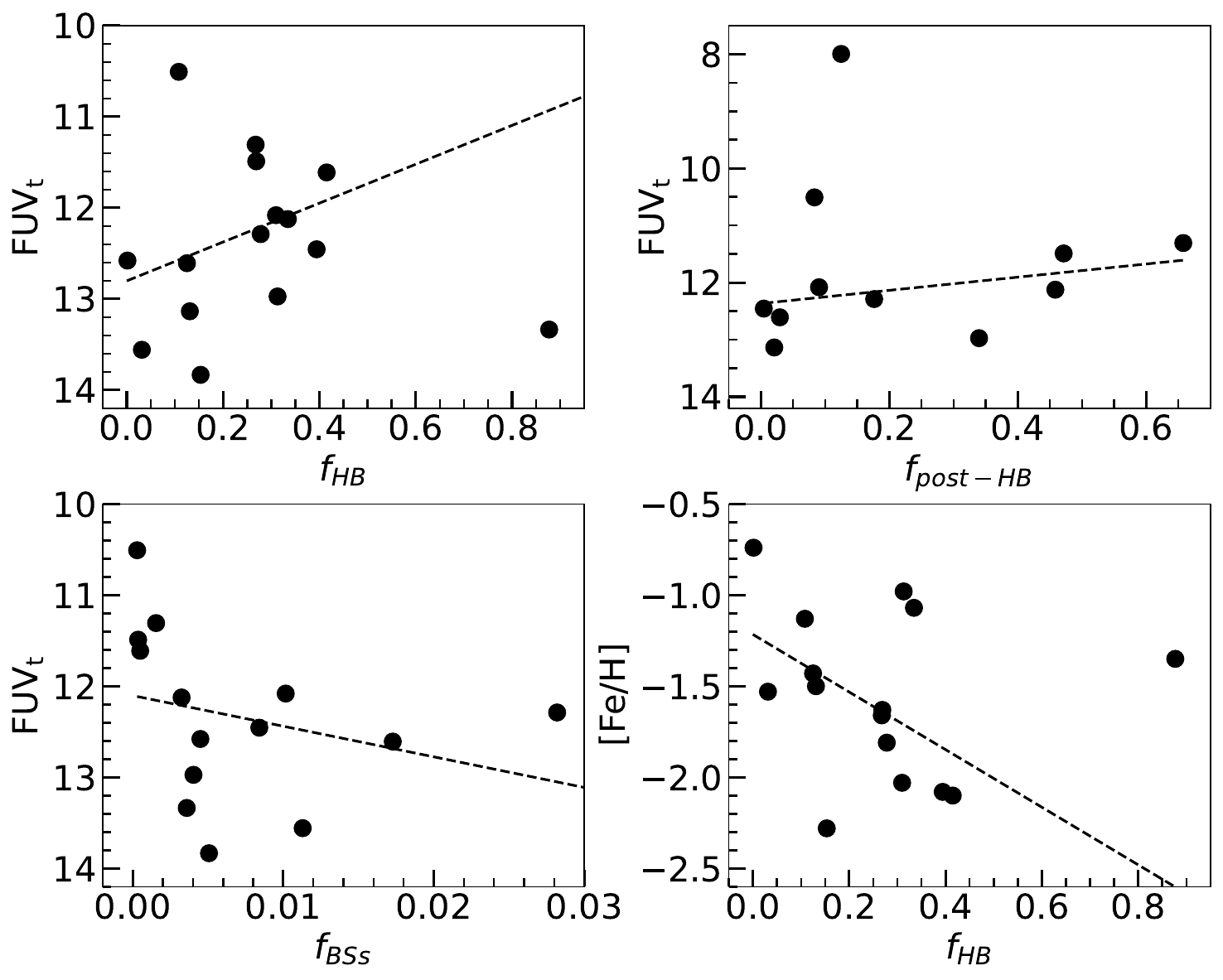} 
          
      \caption{Variation of total integrated FUV emission with fraction of HB (top left panel), post-HB (top right panel), and BS (bottom left panel) emissions. The bottom right panel shows the variation of the fractional contribution of HB stars with metallicity. The dashed line represents the straight-line fitting.}
      \label{fig:frac-ehb}
      \end{figure*}

Most of the stars in the old stellar populations of GCs are too cool to emit significantly in FUV. Instead, the integrated FUV light is likely dominated by blue HB and post-HB stars, with possible small contributions from BSs. Other hot sources, such as X-ray binaries (too rare), white dwarfs, or CVs (too faint), are unlikely to contribute substantially to the total FUV emission \citep{2009catelan_hb, Dalessandro_2012, 2018peacock_hb}. \autoref{tab:gc-source-frac-region} shows the number of different types of evolved stellar sources observed in 18 GCs for which UVIT observations encompass up to the tidal radius ($r_{t}$). We also computed the fraction of surface brightness contribution of each stellar phase to the total surface brightness (given in \autoref{tab:gc-source-frac-region}). This fraction should be used with caution as the sample of evolved stellar population might be incomplete due to unresolved sources in the core of dense clusters and unidentified sources in the cluster due to large distance (e.g., NGC 2419).

A substantial fraction of the total integrated FUV emission in GCs arises from the HB and post-HB stars. Across our sample, post-HB stars contribute up to 45\% to the total integrated FUV light. In GC E3, a single hot blue post-AGB star of magnitude $-2.33$ \citep{2024e3} dominates the UV, contributing 77\% of the total despite the cluster's low mass and the sparse population of such objects. Similarly, in NGC 1904, just two post-HB stars contribute nearly 65\% of the emission, primarily due to one exceptionally bright source with an absolute magnitude of $2.98$. We also highlight that the brightest GC in our sample, NGC 2419 (M$_{\rm FUV} = 8.0$ mag), contains 37 confirmed post-HB stars but no HB stars. This may be due to the large distance of NGC 2419 (the outer halo), as HB stars could not be detected in the FUV images. The next brightest GCs are NGC 2808 (10.51 mag; 39 post-HB stars), NGC 5986 (11.49 mag; 24 post-HB stars), and NGC 1904 (11.31 mag; two post-HB stars). Thus, GCs with larger post-HB populations exhibit systematically higher FUV luminosities. 

 The HB stars contribute up to 40\% of the total FUV emission in most of the GCs. A striking exception is NGC 288, where the complete absence of post-HB stars results in approximately 90\% of the FUV flux originating from its 119 HB stars. Other GCs with large HB populations include NGC 2808 with 198 HB stars, NGC 5986 with 185, NGC 5897 with 160, and NGC 6101 with 148. Across the sample, HB star magnitudes range roughly from $+2$ to $+6$ mag. Despite their large numbers, the collective FUV contribution from HB stars remains lower than that from post-HB stars in clusters where both populations are present. In NGC 2808, despite the large population of HB and post-HB stars, their contribution to the total FUV emission is only about 10\% as these detected sources are intrinsically faint in FUV. Furthermore, the low fractional contribution is likely due to the unresolved HB and post-HB stars in the dense central regions of this GC.

BSs are detected in substantial numbers across our sample of GCs. However, their contribution to the total integrated FUV emission remains minor, accounting for only approximately 3\%.

We present the fractional contributions to the total integrated FUV emission as a function of the cluster’s integrated FUV magnitude. In \autoref{fig:frac-ehb}, we present the fractional contributions of evolved sources to the total integrated FUV emission as a function of the cluster's integrated FUV emission. The top left, top right, and bottom left panels of \autoref{fig:frac-ehb} show the fractional contributions of HB ($f_{HB}$), post-HB ($f_{post-HB}$), and BSs ($f_{BSs}$) stars, respectively. For each population, we fit a linear trend to the data after excluding outliers. Clusters with higher $f_{HB}$ exhibit brighter surface brightness, indicating that the total FUV luminosity increases with the relative HB contribution. A comparable trend is also observed for $f_{post-HB}$: greater post-HB fractional emission corresponds to brighter surface brightness. In contrast, $f_{BSs}$ shows a weak inverse correlation, i.e., higher BS fractions are marginally associated with fainter surface brightness, though the BS contribution is minimal, averaging only $\sim 3$\% across the sample. These results confirm that blue HB and post-HB stars dominate the integrated FUV emission in GCs, driving the overall UV brightness, while BSs play a negligible role.

\subsection{Variation of HB Fractions with Metallicity} \label{sec:HB_fraction_metalicity}

 The fractional FUV emission contributed by HB stars ($\rm f_{HB}$) is plotted as a function of [Fe/H] in the bottom right panel of \autoref{fig:frac-ehb}.  
In metal-poor GCs, HB stars dominate the FUV output. This dominance arises because metal-poor environments favor the formation of hotter, bluer HB stars with effective temperatures exceeding 10,000 K \citep{2025moehler-bhb-abundance}. These hot HB stars are efficient emitters in the FUV wavelength range, where cooler stellar populations contribute negligibly. As a result, $\rm f_{HB}$ has higher values in the most metal-poor clusters, reflecting their significant role in the integrated UV light of these systems. However, as metallicity increases, we observe a clear decline in $f_{\rm HB}$. This trend is driven by changes in HB morphology: higher-metallicity GCs tend to produce cooler, redder HB stars (red HB), which are less effective at generating FUV radiation \citep{2018peacock_hb}. This metallicity-dependent shift has important implications for understanding GC evolution, UV upturn phenomena in elliptical galaxies, and the calibration of stellar population synthesis models, as it highlights how chemical composition influences the late stages of stellar life and the overall SED of ancient stellar systems.

\subsection{Relative distribution of FUV emission across the globular clusters} \label{sec:fuv-distribution}

  The FUV emission across GCs exhibits a wide, nonuniform distribution primarily driven by the presence and population size of hot UV-bright stars \citep{2018peacock_hb}. To investigate the relative distribution of FUV emission across the GCs, we compared the logarithmic ratio of surface brightness in a specific region to that in the tidal region, i.e., $\rm log_{10}(F_{x}/F_{t})$, where $\rm F_{x}$  denotes the area-normalized flux in the core, intermediate, or the outer region, and $\rm F_{t}$ denotes the area-normalized flux in the tidal region.
 
 We found that this fraction is consistently higher in the core region, followed by the intermediate and outer regions as shown in \autoref{tab:gc-source-frac-region}. 
 This dominance solely depends on the central concentration (c) of the GCs. We show the variation of c with the logarithmic ratio of surface brightness in different regions in \autoref{fig:flux-distribution}. We found a strong correlation between surface brightness fraction and the central concentration, i.e., dense clusters have brighter core regions and half-light regions. This is in compliance with the presence of UV-bright HB and post-HB stars in the central regions of the GCs \citep{2021prabhu-ngc2808}. 
 In the outer regions (right panel in \autoref{fig:flux-distribution}), we did not see any such correlation implying the lack of very bright sources in these regions.

 \begin{figure*}
    \centering
     
      \includegraphics[width=\textwidth]{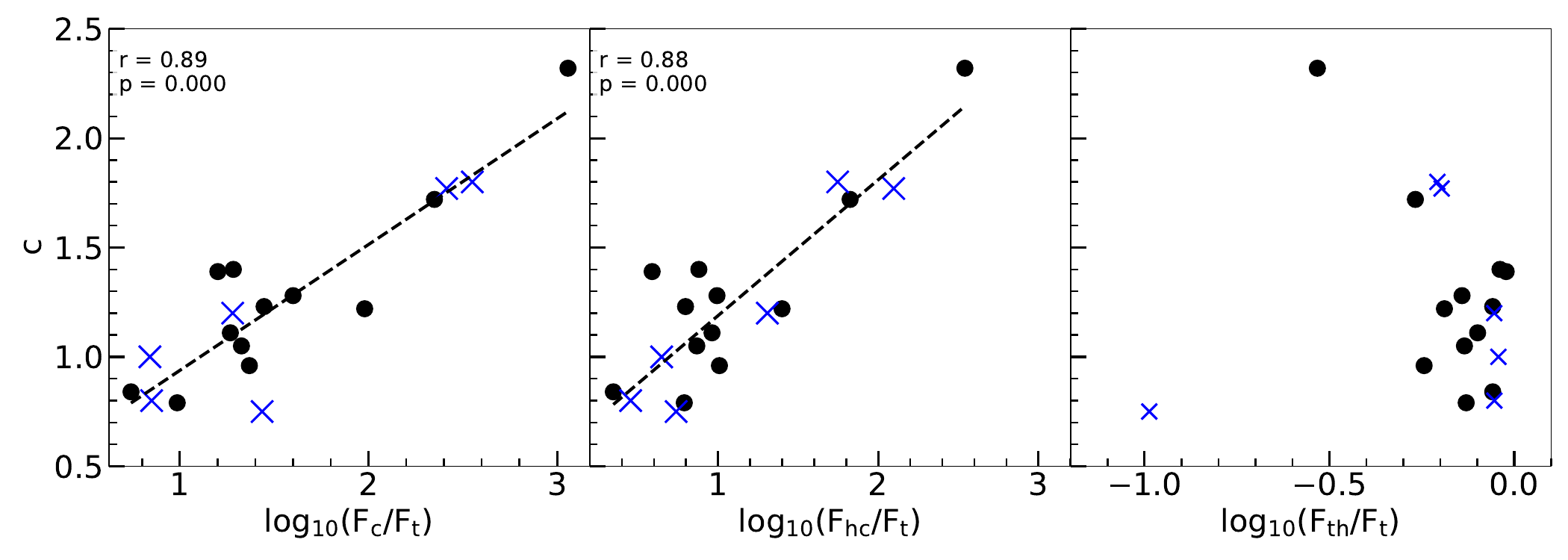} 
          
      \caption{Variation of surface brightness with central concentration parameter. The black dashed line shows the fitting of the data.}
      \label{fig:flux-distribution}
      \end{figure*}
      

We analyzed the contribution of UV-bright evolved sources of the GCs separately across different radial segments (core, intermediate, and outer regions). No clear trend emerges; instead, these sources appear randomly distributed in these regions of the GCs.

\subsection{UV$-$Optical color$-$color diagram} 

\begin{figure}
    \centering
      \includegraphics[width=\columnwidth]{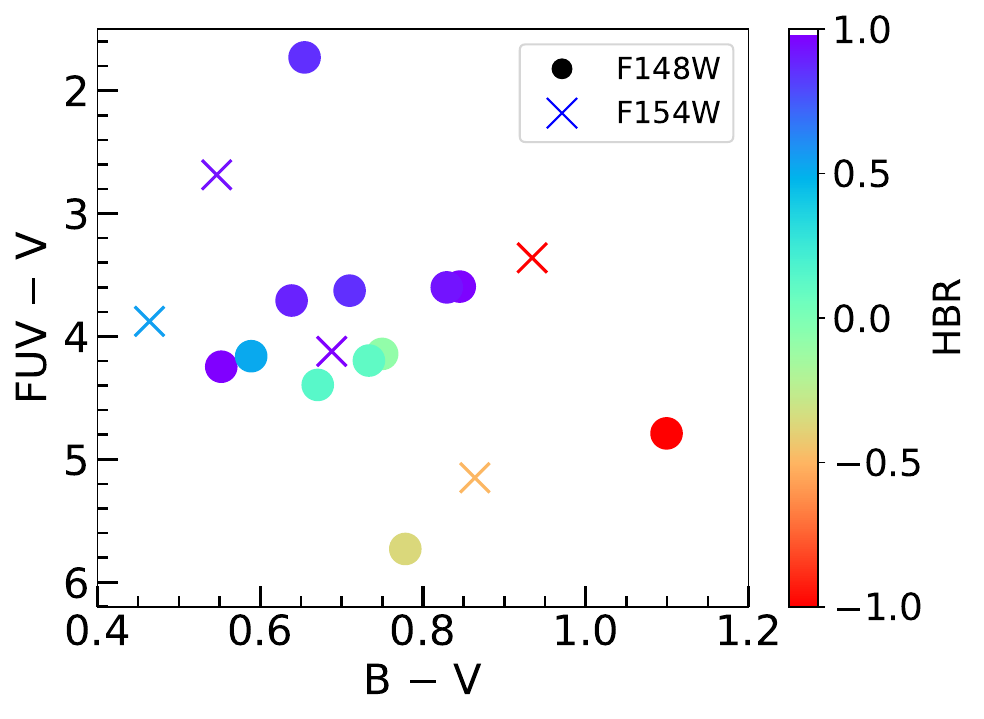}   
      \caption{The B$-$V versus FUV$-$V plot of the GCs. The solid circles and crosses represent the GCs observed in the F148W and F154W filters. The color bar shows the HBR variation taken from \cite{harris2010new} with $\rm HBR=-1$ representing a cluster with a quite red HB morphology and $\rm HBR=1$ representing a cluster with a completely blue HB morphology.}
   
      \label{fig:fuv-optical}
      \end{figure}

FUV$-$optical color spread can also probe the presence of a UV upturn phenomenon in ETGs, compared to the optical color spread taking into account the fractional FUV flux contributions from the hot HB and post-HB stars in them \citep{2007kaviraj, 2011chung-etg, 2018Goudfrooij-uvupturn-gc, 2025two-category-uvupturn}. We compare the integrated optical (B$-$V) colors with the integrated FUV$-$V colors for our observed GCs in \autoref{fig:fuv-optical}. We see that the FUV$-$V color distribution is more dispersed ($\sim$6 mag) than the optical B$-$V color ($\sim$0.6 mag) for our sample. The observed FUV$-$optical colors in the Galactic GCs show a similar spread ($\sim$6 mag) observed in the ETGs with UV-upturn features \citep{1995dorman, 2007kaviraj, 2011chung-etg, 2018Goudfrooij-uvupturn-gc, 2025two-category-uvupturn}.  
The integrated FUV emission in ETGs is not produced by young massive stars (as in star-forming spirals) but arises almost entirely from evolved, low-mass stars of post-MS phases \citep{1993dorman,1995dorman}. The integrated FUV emission predominantly contributed by hot HB and post-HB stars in ETGs \citep{2011chung-etg, 2021ali-sadman} as it is in the case of Galactic GCs \citep{1993dorman, 1995dorman}.

\section{Correlation of cluster parameters with integrated UV magnitudes and UV$-$optical colors} \label{sec:uv-param}

Under this section, we study the effect of cluster parameters on the integrated UV magnitudes and integrated UV$-$optical colors of the GCs.  
\subsection{Integrated UV magnitudes versus cluster parameters} \label{sec:fuv_param}

\begin{figure*}
    \centering
        \includegraphics[width=\textwidth]{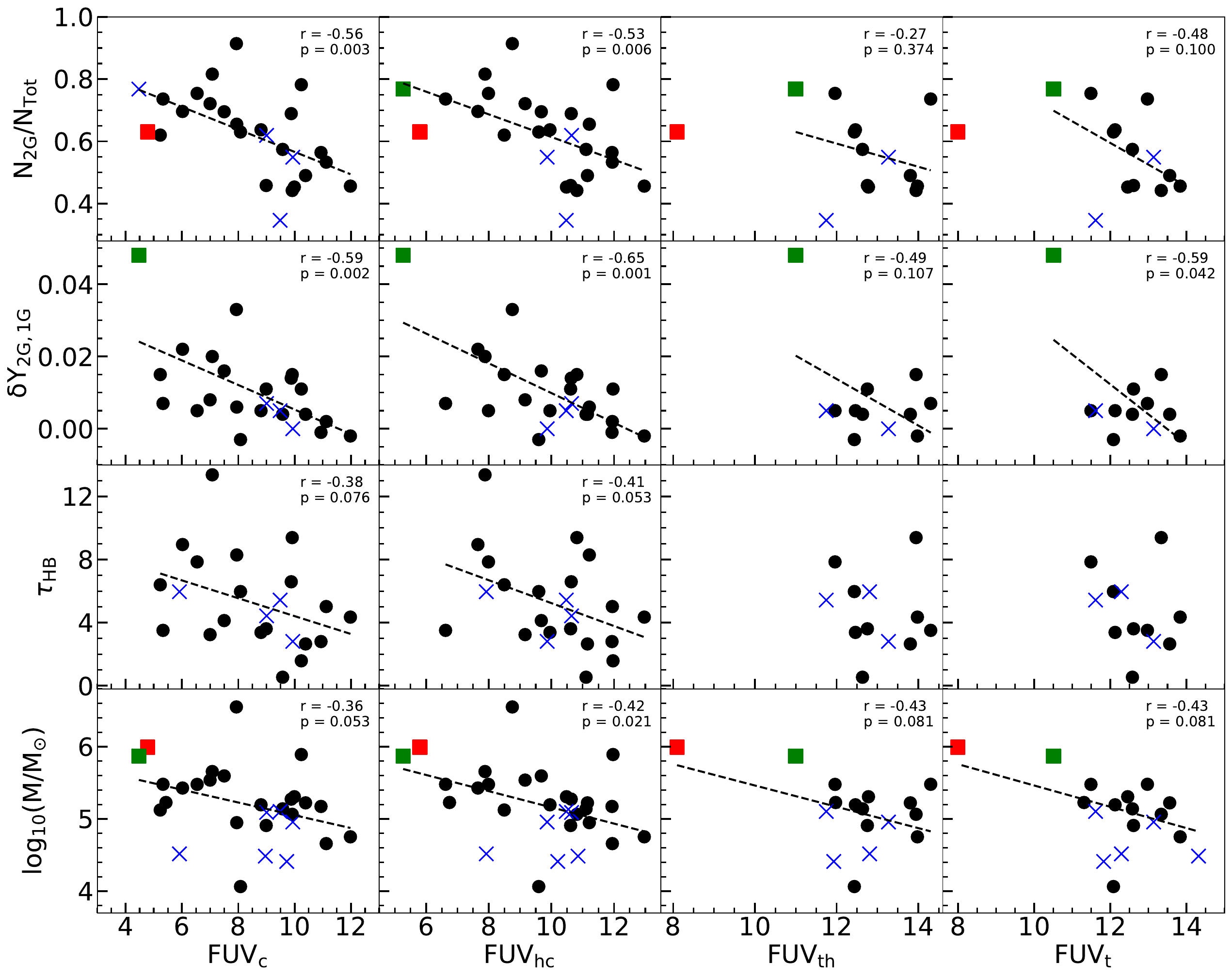} 
\caption{Correlation between surface brightness and cluster parameters for the GCs. The top to bottom panels represent the 2G fraction in a GC ($\rm N_{2G}/N_{Tot}$), the He enhancement in the 2G population with respect to the 1G population ($\delta Y_{2G,1G}$), the HB morphology parameter ($\tau_{HB}$), and the mass of the GC ($\rm log(M/M_{\odot}$), respectively. The blue crosses represent GCs observed in the filter F154W. The red and green square boxes represent NGC 2419 and NGC 2808, respectively. In each panel, the r, p are the Pearson's correlation coefficients and the black dashed line is the fitting line obtained by these coefficients.}
    \label{fig:fuv_param}
\end{figure*}

Almost all GCs host multiple stellar populations: a 1G population with primordial helium abundance (Y $\sim$ $0.23 – 0.25$) and a 2G population enriched in helium ($\Delta Y$ up to $0.10 – 0.15$) along with light elements like N, Na, and Al \citep{2020santi-msp}. This 2G enrichment has a profound impact on the integrated FUV emission of the GC.  The top panels of \autoref{fig:fuv_param} show the variations in surface brightness of the core, intermediate, outer, and overall (up to the tidal radius) regions as a function of 2G fraction, N$_{2G}$/N$_{Tot}$, for the GCs in our sample. In all radial segments, the surface brightness becomes systematically brighter (i.e., lower in magnitude) with increasing $N_{2G}/N_{\rm tot}$, demonstrating a clear and consistent trend throughout the cluster structure. Since FUV light is dominated by hot HB stars, the observed trend is due to the fact that HB stars belonging to the 2G are on an average hotter than those related to 1G, because they have a smaller total mass, and then a thinner envelope that forces them to populate the bluer portion of the HB (see, e.g., \citet{2020santi-msp}, for a detailed discussion on this topic).

 A very similar trend is also observed for the variation of surface brightness with the difference in helium mass fraction between 2G and 1G stars, $\delta Y_{2G,1G}$ of GCs in the core, intermediate, outer, and tidal regions (second row of panels of \autoref{fig:fuv_param}). This enhancement accelerates stellar evolution, reduces envelope mass on the HB, and shifts stars toward hotter effective temperatures. This is consistent with \cite{2015Piotto-msp}, who report that GCs host multiple populations and that the GCs with high N$_{2G}$/N$_{Tot}$ are systematically brighter. One of the GCs, NGC 2808 ($\rm FUV_{t} = 10.51$ mag), with the maximum $\delta Y_{2G,1G}$, is brighter than all of the other GCs and has a significant number of HB and post-HB stars, implying He-enhanced and brighter stars in this GC.

\cite{2019new-hbr} introduced a new HB morphology parameter, $\tau_{HB}$, defined as the ratio of areas under the cumulative number distribution of star counts in magnitude (I) and color (V$-$I) for 60 GCs, using both ground- and space-based data. They claimed that $\tau_{HB}$ is much more robust than the previous HB morphology parameters defined in the literature. This parameter shows a positive correlation with cluster age and an anticorrelation with metallicity. In the third row of panels of \autoref{fig:fuv_param}, we show the relationships between the $\tau_{HB}$ and the surface brightness in the core, intermediate, and outer regions. In both the core and intermediate regions, $\tau_{HB}$ increases with FUV emission in a linear correlation. This trend supports the fact that the brighter GCs possess bluer HB morphologies and larger spreads in the magnitude distribution. 
No clear trend emerges in the outer regions, possibly due to the limited number of clusters in this region.

 \begin{figure*}
    \centering
        \includegraphics[width=\textwidth]{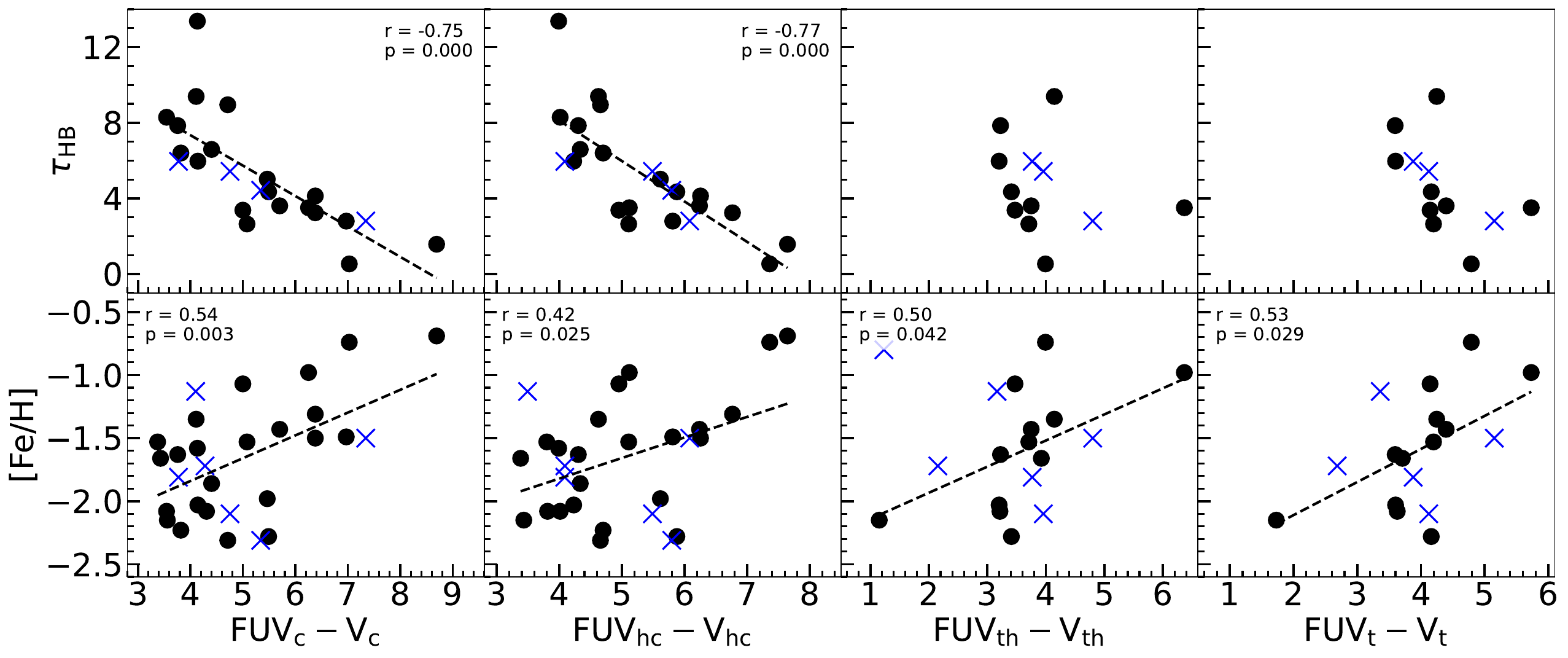} \\ 
    
\caption{Correlation between integrated UV$-$optical colors and cluster parameters for GCs. The upper and lower panels represent the HB morphology parameter ($\tau_{HB}$) and metallicity ([Fe/H]), respectively. 
In each panel, the r, p are the Pearson's correlation coefficients  and the black dashed line is the fitting line obtained by these coefficients.}
    \label{fig:fuv_v_param}
\end{figure*}

The fourth row of panels in \autoref{fig:fuv_param} shows the variation of surface brightness with $\rm log_{10}(M/M_{\odot})$ of the GCs in the core, intermediate, outer, and tidal regions. In all the regions, the brighter GCs are massive, and the mass of the GCs decreases as the GCs move toward the fainter end. This trend agrees with \cite{2015Piotto-msp} who state that the GCs host multiple populations and the GCs with high N$_{2G}$/N$_{Tot}$ are brighter in the UV. This is primarily because of the fact that more massive GCs host a larger fraction of 2G stars, which are helium-enhanced ($Y \gtrsim 0.30$) and evolve into hotter HB and post-HB stars. These stars dominate the FUV emission of GCs, contributing more than 80\% of the total integrated FUV emission despite comprising only a small fraction of the stellar mass. 

We have compared our surface brightness with other parameters such as age, [Fe/H], and HBR, but we did not find any significant correlation between surface brightness and these parameters in any region.

\subsection{Integrated UV$-$optical colors versus cluster parameters}

GCs exhibit a wide range of HB morphologies, which significantly influence their integrated UV-to-optical colors. Bluer HB morphologies (higher HBRs) feature more hot, UV-bright stars, which dominate the far-UV flux and make integrated colors bluer \citep{Dalessandro_2012}. The upper panels of \autoref{fig:fuv_v_param} illustrate the variation of FUV$-$V color with $\tau_{HB}$ across the core, intermediate, and outer regions of GCs. In both the core and intermediate regions, the bluer GCs have higher $\tau_{HB}$, indicating the presence of hotter HB stars. As the integrated FUV$-$V color becomes redder, $\tau_{HB}$ systematically decreases, confirming that bluer and UV-brighter GCs possess blue HB morphologies. 

Integrated UV$–$optical colors of GCs are strongly influenced by metallicity [Fe/H], primarily through its effect on HB morphology, stellar opacity, and the emergence of hot UV-bright stars. The lower panels of \autoref{fig:fuv_v_param} show the variations of integrated FUV$-$V colors with the [Fe/H] across the core, intermediate, and outer regions of GCs. In the core and intermediate regions, (FUV$-$V) colors exhibit significant scatter with [Fe/H], reflecting the dominant influence of HB morphology and localized hot stellar populations rather than a simple metallicity trend. By contrast, in the outer and tidal regions, a clearer trend emerges: FUV magnitudes become progressively brighter (i.e., bluer (FUV$-$V) colors) as metallicity decreases from metal-rich to metal-poor regimes. This behavior aligns with expectations from optical colors, where metal-poor GCs are systematically bluer and metal-rich GCs are redder due to cooler stellar envelopes at lower [Fe/H] \citep{2024metal-rich-poor}. Thus, while inner regions are dominated by stochastic UV contributions from hot HB and post-HB stars leading to scatter, the outer regions reveal the underlying metallicity-driven trend in UV–optical colors, consistent with population synthesis predictions for old, metal-poor stellar systems.

We also compared the total tidal UVIT integrated FUV$-$V colors of the GCs with the GALEX observations \citep{Dalessandro_2012} for  [Fe/H]. The total GALEX integrated FUV$-$V color is dominantly redder for metal-rich GCs ([Fe/H] $>$ $-$1), varies $\sim$4 mag for the GCs having $-$1.5 $\le$ [Fe/H] $<$ $-$1, and blue for GCs with [Fe/H] $\le$ $-$1.5. The spread in the FUV$-$V colors is attributed to a ``second parameter'' effect. In UVIT, we observe similar redder FUV$-$V colors for the metal-rich clusters with [Fe/H] $>$ $-$1. But for the GCs with $-$2.0 $\le$ [Fe/H] $<$ $-$1, the FUV$-$V colors have a spread of $\sim$2 mag (as shown in the bottom rightmost panel of \autoref{fig:fuv_v_param}).

We have compared our integrated colors with other parameters, such as the second-generation fraction, helium mass fraction, mass, central concentration, age, [Fe/H], and HBR values, but we did not find any significant correlation between our FUV colors and these parameters in any region.

\section{Summary}
We have presented integrated FUV magnitudes and colors for 30 Galactic GCs, derived from AstroSat/UVIT images in the F148W and F154W filters. Measurements were performed in the three spatial regions, i.e., core, intermediate, and outer regions, for each cluster. The sources of various evolutionary stages (mainly HB, BS, and post-HB stars) were identified using FUV$-$optical vs. FUV CMDs and BaSTI-IAC isochrones. Cluster parameters (metallicity, age, distance, total mass, 2G fraction, and HBR) were adopted from the most recent literature.  Surface brightness was measured by aperture photometry, corrected for foreground field-star contamination and extinction, and converted to magnitudes in the AB system. Region-specific surface brightness was provided, along with total surface brightness up to the tidal radius ($\rm FUV_{t}$) for the 18 clusters whose tidal radii lie within the UVIT FoV. Complementary optical (V-band) surface brightness was derived from the HUGS catalog (inner regions) and GBT data (outer regions).  

 We found the 100\% completeness limits in a GC to be 22.8 mag for the intermediate and tidal regions and 50\% completeness limits to be 23.5 mag and 23 mag in the intermediate and tidal regions, respectively.

 We quantified the fractional contribution of HB, Post-HB, and BS stars to the total FUV light of each cluster. The post-HB stars contribute up to 45\% to the total integrated FUV light. We found a linear positive correlation between the fractional contribution of surface brightness due to post-HB stars and the total surface brightness of each cluster, i.e.,  the greater the post-HB fractional flux, the brighter the FUV surface brightness. The HB stars contribute up to 40\% of the total FUV emission. HB stars also show a linear positive correlation between the HB surface brightness and the total surface brightness. The contribution of BSs to the total integrated FUV emission is approximately 3\%. We found a weak inverse correlation between the fractional flux of BSs and the total surface brightness.

The UV upturn phenomenon in ETGs and Galactic GCs is mostly dominated by hot HB and post-HB stars. ETGs with UV upturn show a spread in FUV$-$optical colors and we observed a similar spread in FUV$-$optical colors of the sampled Galactic GCs. This may help us to explore FUV$-$optical versus optical color$-$color diagrams of extragalactic GCs while exploring the UV upturn in other galaxies.  

 We found a systematic increase in the logarithmic ratio of surface brightness in the core and intermediate regions to the total surface brightness in the tidal region. The FUV emission in both the core and intermediate regions strongly depends on the central concentration parameter of the cluster.

 We studied the correlation between the surface brightness and the cluster parameters such as the $\rm N_{2G}/N_{Tot}$, $\rm \delta Y_{2G,1G}$, $\rm \tau_{HB}$, mass and [Fe/H] of the GCs in the core, intermediate, outer, and tidal regions. We found that the surface brightness, which is dominated by the hotter and bluer HB stars, becomes brighter with an increase in $N_{2G}/N_{Tot}$ and $\rm \delta Y_{2G,1G}$ values in all the regions. Since the massive GCs host a larger fraction of helium-enhanced 2G stars, which evolve into hot HB and post-HB stars, they show relatively brighter surface brightness in all the cluster regions. The GCs with high $\rm \tau_{HB}$ parameters have brighter surface brightness and bluer FUV$-$V colors in the core and intermediate regions, confirming bluer HB morphologies in these regions. No clear trend is observed in the outer regions. The FUV$-$V colors are scattered with [Fe/H] in the core and intermediate regions, but in the tidal regions, we found a clear trend between the FUV$-$V colors and [Fe/H] values of GCs. The FUV$-$V colors become redder to bluer as the [Fe/H] decreases from metal-rich to metal-poor. 

 Overall, we have provided the surface brightness values of sample of 30 GCs and an insight into their correlation with several cluster parameters. Such correlations are critical for deciphering the nature of distant systems, which only present integrated properties. This database of surface brightness of GCs is essential for advancing our understanding of extragalactic stellar populations. Moreover, direct comparisons of the integrated UV properties of Galactic and extragalactic GCs  will provide invaluable insights into the stellar population content of these systems. 

\begin{acknowledgments}
We thank Prof. Santi Cassisi for providing valuable suggestions while we were preparing the draft. This publication uses the data from the AstroSat mission of the ISRO, archived at the Indian Space Science Data Center (ISSDC). A.C.P. and S.P. acknowledge the support of the Indian Space Research Organisation (ISRO) under the AstroSat archival data utilization program (No. DS\_2B-13013(2)/1/2022-Sec.2). A.C.P. also thanks the Inter University Center for Astronomy and Astrophysics (IUCAA), Pune, India, for providing facilities to carry out his work. D.P. acknowledges funding from the HTM (grant TK202), ETaG (grant PRG 3034), and EU Horizon Europe (EXCOSM, grant No. 101159513).
\end{acknowledgments}

%



\appendix
\section{Estimation of the UV Sky background for faint UV sources within a cluster}
Background for a UVIT image is not constant across the FoV. In \autoref{fig:bkg-var},  we show the variation of a background map across a single GC NGC 6723 (left panel) and the background variation with distance from the center of the cluster up to its tidal radius (middle panel). The mean background is higher in the central region up to the half-light radius where the crowding is higher. We estimated the background using the segmentation maps obtained from source-extractor after masking the sources above the 2$\sigma$ threshold. Any source below this threshold was considered as background. In the right panel of \autoref{fig:bkg-var}, we show the surface brightness plot computed within concentric circles in units of core radius for the raw data (blue triangles) and the data after sky subtraction (black solid circles). The flattening of the raw data shows that the light from faint stars (for e.g. red giant branch and MS stars which emit flux dominantly in the optical bands) approaches the background magnitude. Even in the sky-subtracted data, the flux is very low beyond the background limit (horizontal dashed line). 

\begin{figure}
    \centering

       \includegraphics[width=0.33\linewidth]{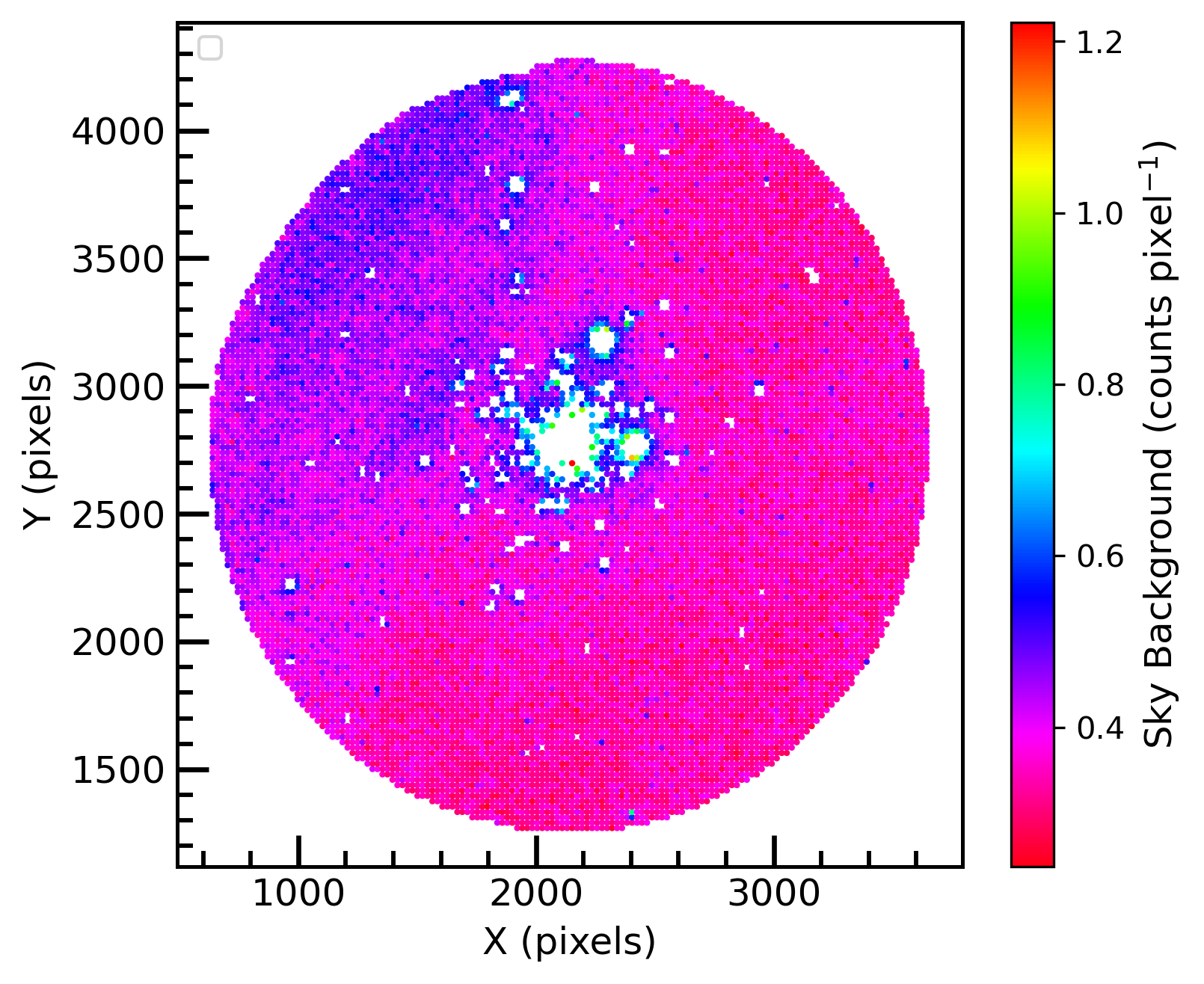}
           \includegraphics[width=0.31\linewidth]{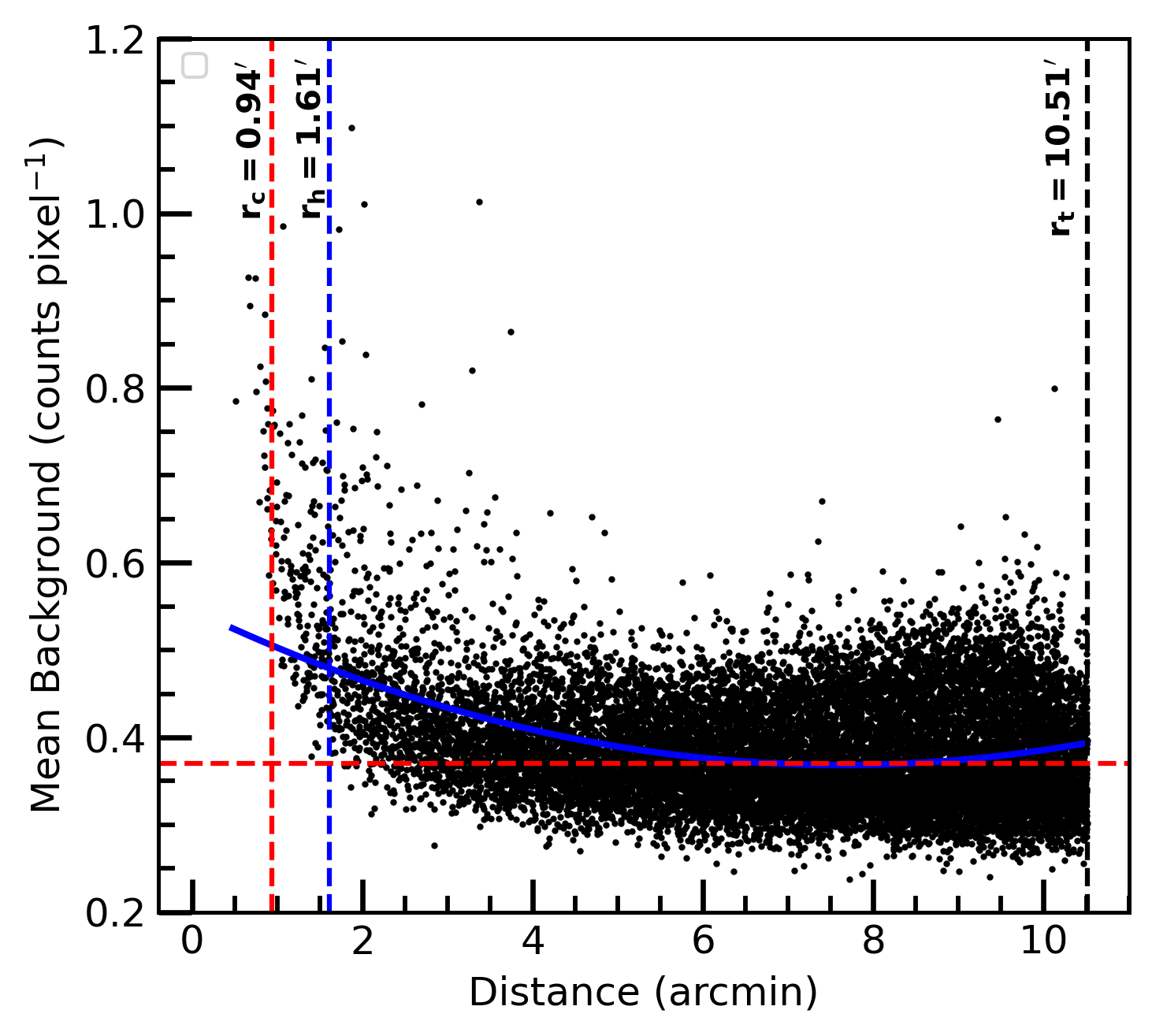}
          \includegraphics[width=0.33\linewidth]{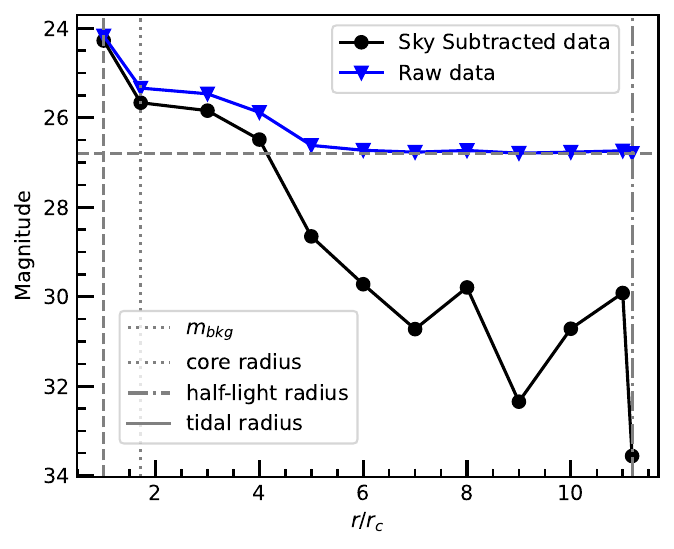}
    \caption{Left panel: we show the variation of the background map across the image in the color map.  Middle panel: variation of background with the distance of the source from the center. The red dashed horizontal line is the mean background we have calculated for the image (0.37 counts pixel$\rm ^{-1}$). We subtracted this background from the source flux. Right panel: we show the surface brightness profile versus radius (in units of core radius: 0.94$'$).}
    \label{fig:bkg-var}
\end{figure}


\bibliography{ref}{}
\bibliographystyle{aasjournalv7}



\end{document}